\documentclass[iop,apj]{./emulateapj}
\usepackage{epsfig}
\usepackage{color}

\newcommand   {\about} {\mbox{$\sim$}}
\newcommand   {\mic}   {\mbox{$\mu$m}}

\newcommand   {\htop}  {\mbox{H$_3$O$^+$}}

\newcommand   {\nhhh}  {\mbox{NH$_3$}}

\newcommand   {\arcs}  {\mbox{$^{\prime\prime}$}}

\newcommand   {\pscm}  {\mbox{cm$^{-2}$}}
\newcommand   {\kms}   {\mbox{km\,s$^{-1}$}}

\renewcommand {\ga}    {\mbox{\rlap{\hbox{\lower5pt\hbox{$\sim$}}}\hbox{$>$}}}
\renewcommand {\la}    {\mbox{\rlap{\hbox{\lower5pt\hbox{$\sim$}}}\hbox{$<$}}}

\received{2014 January 16}
\accepted{2014 March 4}

\slugcomment{To appear in The Astrophysical Journal}

\shortauthors{Lis et al.}

\shorttitle{Widespread Hot Hydronium Ion}

\begin{document}

\title{Widespread Rotationally-Hot Hydronium Ion in the Galactic Interstellar Medium}

\author{D.~C.~Lis$^{1,2}$, P.~Schilke$^3$, E.~A.~Bergin$^4$,
  M.~Gerin$^5$, J.~H.~Black$^6$, C.~Comito$^3$, M.~De~Luca$^5$,
  B.~Godard$^5$, R.~Higgins$^3$, F.~Le~Petit$^5$, J.~C.~Pearson$^7$,
  E.~W.~Pellegrini$^8$, T.~G.~Phillips$^1$, and S.~Yu$^7$}

\altaffiltext{1}{California Institute of Technology, Cahill Center for
  Astronomy and Astrophysics 301-17, Pasadena, CA~91125, USA; 
  dcl@caltech.edu, tgp@submm.caltech.edu}

\altaffiltext{2}{Sorbonne Universit\'{e}s, Universit\'{e} Pierre et
  Marie Curie, Paris 6, CNRS, Observatoire de Paris, UMR 8112, LERMA,
  Paris, France}

\altaffiltext{3}{I. Physikalisches Institut, University of Cologne,
  Z\"{u}lpicher Str. 77, 50937 K\"{o}ln, Germany;
  schilke@ph1.uni-koeln.de, ccomito@ph1.uni-koeln.de,
  higgins@ph1.uni-koeln.de} 

\altaffiltext{4}{University of Michigan, Ann Arbor, Michigan 48109,
  USA; ebergin@umich.edu}

\altaffiltext{5}{\'{E}cole Normale Superi\'{e}ure, CNRS, Observatoire
  de Paris, UMR 8112, LERMA, Paris, France;
  maryvonne.gerin@lra.ens.fr, deluca@lra.ens.fr,
  benjamin.godard@lra.ens.fr, franck.lepetit@obspm.fr}

\altaffiltext{6}{Department of Earth and Space Sciences, Chalmers
  University of Technology, Onsala Space Observatory, SE-43992 Onsala,
  Sweden; john.black@chalmers.se}

\altaffiltext{7}{Jet Propulsion Laboratory, California Institute of
  Technology, 4800 Oak Grove Drive, Pasadena, CA 91109, USA;
  john.c.pearson@jpl.nasa.gov, shanshan.yu@jpl.nasa.gov} 

\altaffiltext{8}{Department of Physics and Astronomy, University of
  Toledo, Toledo, OH 43606, USA; eric.pellegrini@utoledo.edu} 

\begin{abstract}
  We present new \emph{Herschel}\footnote{Herschel is an ESA space
    observatory with science instruments provided by European-led
    Principal Investigator consortia and with important participation
    from NASA.} observations of the (6,6)\footnote{The correct
    spectroscopic notation should be $J_K = 6_6^- \leftarrow 6_6^+$,
    but we follow here the notation commonly used in the ammonia
    literature.} and (9,9) inversion transitions of the hydronium ion
  toward Sagittarius B2(N) and W31C. Sensitive observations toward
  Sagittarius~B2(N) show that the high, \about 500~K, rotational
  temperatures characterizing the population of the highly-excited
  metastable \htop\ rotational levels are present over a wide range of
  velocities corresponding to the Sagittarius~B2 envelope, as well as
  the foreground gas clouds between the Sun and the source.
  Observations of the same lines toward W31C, a line of sight that
  does not intersect the Central Molecular Zone, but instead traces
  quiescent gas in the Galactic disk, also imply a high rotational
  temperature of \about 380~K, well in excess of the kinetic
  temperature of the diffuse Galactic interstellar medium. While it is
  plausible that some fraction of the molecular gas may be heated to
  such high temperatures in the active environment of the Galactic
  center, characterized by high X-ray and cosmic ray fluxes, shocks
  and high degree of turbulence, this is unlikely in the largely
  quiescent environment of the Galactic disk clouds. We suggest
  instead that the highly-excited states of the hydronium ion are
  populated mainly by exoergic chemical formation processes and
  temperature describing the rotational level population does not
  represent the physical temperature of the medium. The same arguments
  may be applicable to other symmetric top rotors, such as ammonia.
  This offers a simple explanation to the long-standing puzzle of the
  presence of a pervasive, hot molecular gas component in the central
  region of the Milky Way. Moreover, our observations suggest that
  this is a universal process, not limited to the active environments
  associated with galactic nuclei. \vspace {5pt}
\end{abstract}


\section{Introduction}

The presence of molecules in interstellar space was predicted as early
as 1930's (e.g., \citealt{swings37}), quickly followed by
identification of simple diatomic species, CH, CN, CH$^+$ (e.g.,
\citealt{mckellar40}), in optical absorption spectra. These early
observations lead to important insights into the physics and chemistry
of the interstellar medium (e.g., \citealt{bates51}). However, the
optical and UV observations are limited to diffuse or translucent
clouds and applications of molecular spectroscopy to studies of the
dense interstellar medium had to await the development of microwave
detection techniques in the 1960s. Since the first microwave
detections of molecules in space \citep{weinreb63,cheung69,wilson70},
molecular spectroscopy has been a premier tool for studying
interstellar gas cloud physics, from the Milky Way \citep{evans99} to
the high-redshift universe \citep{solomon05}. In order to use
molecules as quantitative tracers of star-forming clouds, it is
necessary to understand the radiative and collisional processes that
govern the excitation of their internal quantum states
\citep{flower07}. For decades, it has been theorized that, under
interstellar conditions, some molecules form via highly-exothermic
reactions between ions and neutral molecules \citep{herbst73}.
However, quantitative models of chemistry and of excitation have
conventionally been constructed independently, on the grounds that
chemical time-scales are long compared with excitation time-scales.

The most abundant interstellar molecule, molecular hydrogen, is a
symmetric rotor without a permanent dipole moment, which only has weak
quadrupole rovibrational transitions in its ground electronic state
(e.g., \citealt{habart05}). Its UV electronic transitions can only be
studied from space and the observations are limited to the relatively
nearby diffuse and translucent clouds. The infrared rovibrational
lines are accessible from the ground, but are only excited in warm,
active environments, such as shocks or photon dominated regions.
Consequently, trace molecules, with rich rotational spectra in the
millimeter to far-infrared wavelength range have been used as proxies
of H$_2$, to understand the physical conditions in the star-forming
gas reservoir of the Milky Way and external galaxies. Of these, oblate
symmetric top molecules, in particular ammonia \citep{ho83}, for which
the $J=K$ level is the lowest energy level in each $K$ ladder, are of
great utility for the temperature determination. The population of
these ``metastable'' levels is assumed to be thermalized at the
kinetic temperature of the medium, due to the absence of allowed
radiative transitions out of these levels.

The hydronium ion, \htop, is another oblate symmetric rotor,
isoelectronic with ammonia, which like ammonia has the characteristic
inversion splitting of its rotational levels. In the case of \htop,
however, the splitting is very large \citep{liu85} and the inversion
transitions occur at far-infrared wavelengths, as opposed to
centimeter wavelengths for ammonia. The far-infrared \htop\ inversion
transitions, up to (11,11), have only recently been detected for the
first time by \emph{Herschel}, in absorption toward the central region
of the Milky Way \citep{lis12}.

Earlier HIFI observations of the hydronium ion toward Sagittarius~B2
indicated a high rotational temperature, \about 500~K, at velocities
corresponding to the cloud envelope. \cite{lis12} suggested that, in
regions exposed to ionizing radiation, such as those at the center of
the Milky Way, highly-excited states of the hydronium ion and ammonia
are populated mainly by exoergic chemical formation processes and
temperatures derived from their spectra do not represent the physical
temperature of the medium. The detection of highly-excited states of
the hydronium ion in nearby active galaxies, with rotational
temperatures similar to that observed in Sagittarius B2
\citep{gonzalez13}, demonstrated the universality of the shape of the
observed population diagrams in \emph{active} environments. Here we
present new sensitive \emph{Herschel} observations of the hydronium
ion on the lines of sight toward Sagittarius B2(N) and W31C
(G10.6-0.4), which clearly demonstrate that this effect is not limited
to the environments of galactic nuclei, characterized by high X-ray
and cosmic ray fluxes, but is a \emph{fundamental} property, pervasive
throughout the interstellar medium.

\section{Observations}

The inversion lines of \htop\ were first discovered in absorption in
the full spectral scans of Sagittarius~B2(M)\footnote{\emph{Herschel}
  OBSIDs: 1342191565, 2546, 2656, 1342204723, 4739, 5848, 6455, 6501,
  664, 1342215935, 6702, and 8200; OD 292--693.} and
(N)\footnote{\emph{Herschel} OBSIDs: 1342204692, 4703, 4731, 4812,
  4829, 5491, 5855, 6364, 6370, 6498, 6643, 1342215934, 6701, and
  8198; OD 489--693.} carried out between 2010 March and 2011 April
using the Heterodyne Instrument for the Far-Infrared (HIFI;
\citealt{degraauw10}) on the Herschel Space Observatory
\citep{pilbratt10}, within the framework of the HEXOS guaranteed time
key program \citep{bergin10}. Preliminary analysis of the
Sagittarius~B2(N) data was presented by \cite{lis12}. The data
presented here have been re-reduced using an improved version of the
HIFI pipeline, which results in significantly lower noise levels in
the high-frequency HEB mixer bands. The source coordinates are:
Sagittarius~B2(M) $\alpha_{J2000}=17^{\rm h}47^{\rm m}20.35^{\rm s},
\delta_{J2000}=-28^{\circ}23^{\prime}03.00^{\prime\prime}$;
Sagittarius~B2(N) $\alpha_{J2000}=17^{\rm h}47^{\rm m}19.88^{\rm s},
\delta_{J2000}=-28^{\circ} 22^{\prime}18.40^{\prime\prime}$.

The line survey data have been calibrated with HIPE version 10.0
\citep{roelfsema12} and the resulting double-sideband (DSB) spectra
were subsequently reduced using the GILDAS
CLASS\footnote{http://www.iram.fr/IRAMFR/GILDAS} software package.
Basic data reduction steps included removal of spurious features or
otherwise unusable portions of the spectra. The continuum emission was
then subtracted from the DSB scans by fitting a low-order (typically
0--1, in a few cases 2) polynomial. The continuum-subtracted DSB data
were deconvolved (sideband separation through pure $\chi^2$
minimization; \citealt{comito02}) to provide a single-sideband (SSB)
spectrum for each HIFI band.

The variation of the continuum intensity as a function of frequency
was determined from the least-squares fit to the emission-free
spectral channels across all HIFI bands. The fitted continuum was
subsequently folded back into the deconvolved, baseline subtracted
spectra. The data reduction procedure ensures that saturated
absorption features reach the zero continuum level and there are no
discontinuities in the overlap regions between the various HIFI bands.

In addition to the full HIFI spectral scans, we obtained deep
follow-up integrations at the frequencies of the (6,6) and (9,9)
inversion transitions of the hydronium ion toward Sagittarius B2(N)
and W31C. These observations\footnote{\emph{Herschel} OBSIDs:
  1342252167--169, 174--176, 134253608, 610--12; OD 1239 and 1252.}
were carried out in 2012 October 4--17 in the framework of the
special, final call for \emph{Herschel} observing proposals, referred
to by ESA as ``Must-Do'' programs. The source coordinates for W31C are
$\alpha_{J2000}=18^{\rm h}10^{\rm m}28.70^{\rm s},
\delta_{J2000}=-19^{\circ}55^{\prime}50.00^{\prime\prime}$ and
  for Sagittarius~B2 are the same as the earlier HEXOS observations.
The data were taken in the double beam switching (DBS) mode, with the
reference beams \about 180\arcs\ away from the source, using the HIFI
wideband spectrometer (WBS), which provides a 1.1.~MHz frequency
resolution over the 2.4~GHz bandwidth of the band 7 HEB receivers.
Three shifted LO settings were observed, which were subsequently
averaged together after inspecting the spectra for possible line
contamination from image sideband (no contamination was found). The
total observing time for each of the two sources was 53 and 56 min for
the (6,6) and (9,9) lines, respectively.

Spectra from the HIFI HEB mixer bands are known to be adversely
affected by standing waves originating in the signal amplification
chain. This problem is particularly apparent in long integrations,
such as the deep Sagittarius~B2(N) and W31C observations presented
here, where the detector power levels can drift significantly,
resulting in a disparity between the ``on'' and ``off'' phases of the
observation. Since the standing waves occur in the electrical chain,
where the reflecting components have complex frequency-dependent
impedance, the resulting baseline ripple has a non-sinusoidal behavior
and it is difficult to remove using standard standing wave removal
techniques \citep{higgins09}. Fortunately the standing wave phase and
amplitude are stable for a given mismatch between the ``on'' and ``off''
phases. Therefore, the spectra can be corrected by subtracting a
standing wave profile from similarly mismatched template spectra. This
approach has been applied here to the individual integrations, before
they were averaged together to produce the final spectra.

\begin{figure*}[t]
\centering

\includegraphics[width=0.85\textwidth,angle=0]{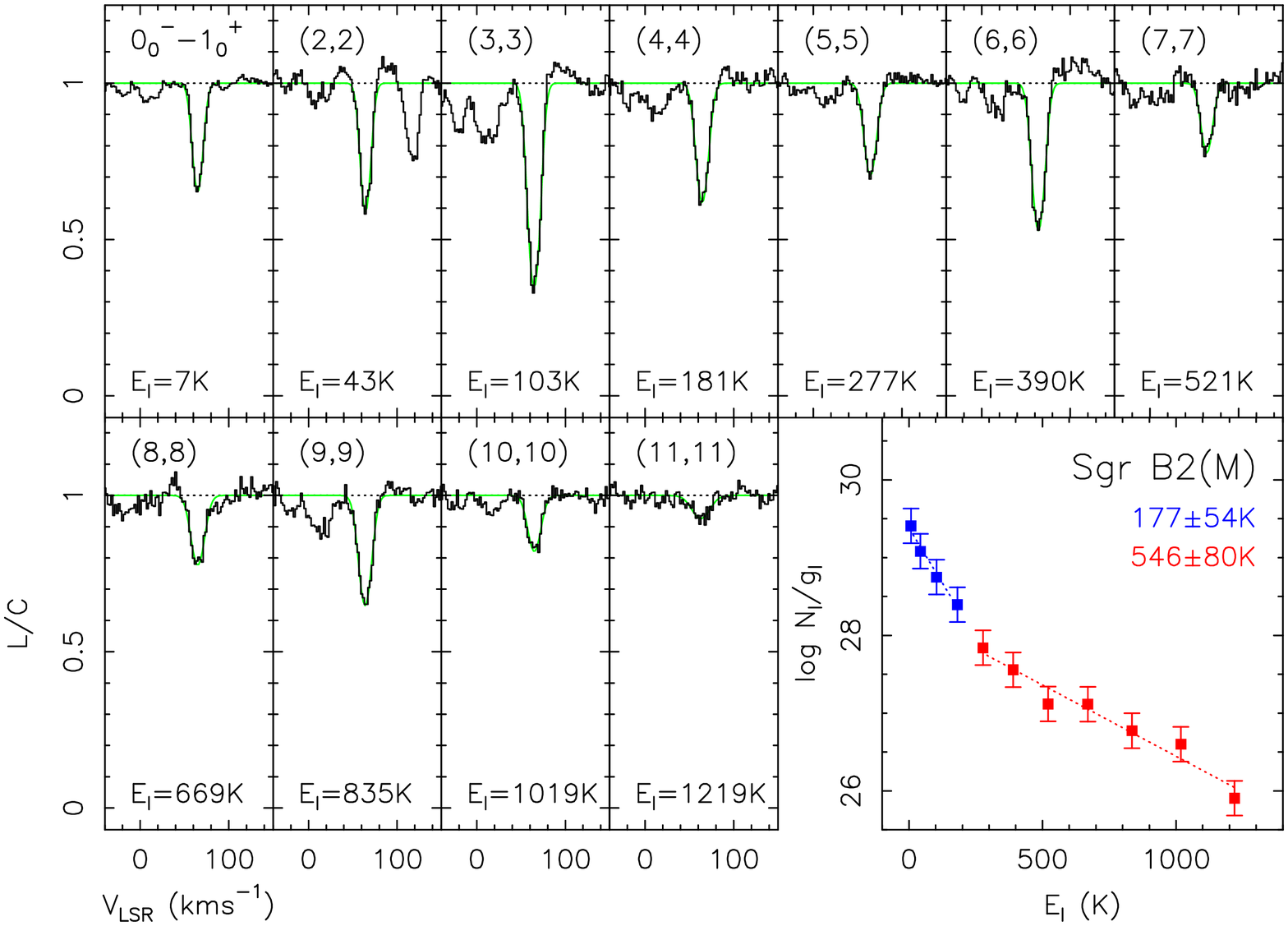} 
\includegraphics[width=0.85\textwidth,angle=0]{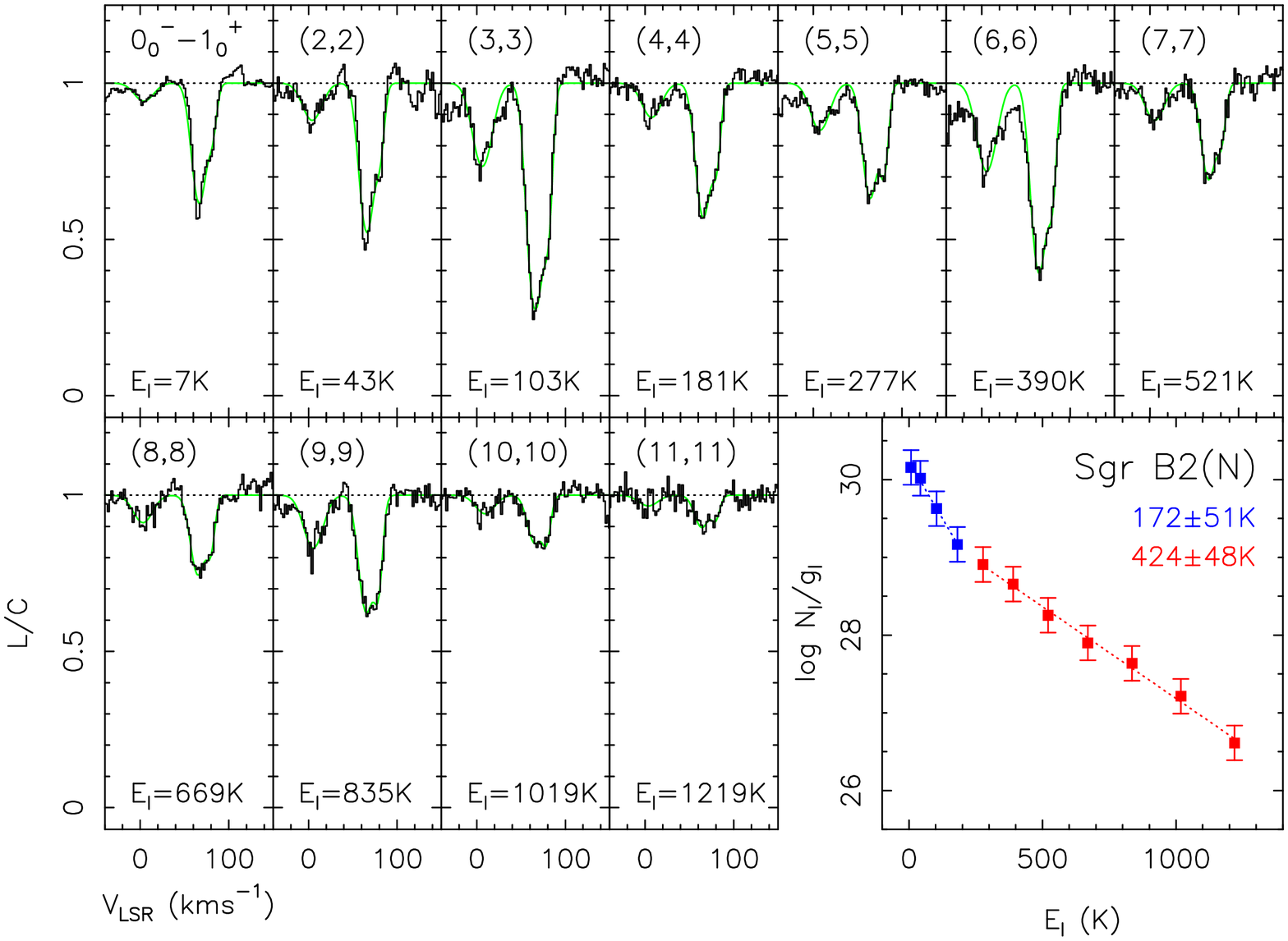}
\caption{Spectra of the metastable inversion transitions of \htop\
  toward Sagittarius~B2(M) and (N), upper and lower panels,
  respectively, and the corresponding rotation diagrams. The rotation
  diagram for Sagittarius~B2(M) corresponds to the 64~\kms\ component,
  associated with the cloud envelope, while for Sagittarius~B2(N) it
  corresponds to the sum of the three absorption components seen in
  the spectra, at \about 5, 65, and 80~\kms. Green lines are Gaussian
  fits to the spectra, three components for Sagittarius B2(N) and one
  component for Sagittarius B2(M). The \emph{maximum} uncertainties of
  the individual measurements (dominated by systematics) are
  conservatively assumed to be 25\%. The rotational temperatures of
  the warm and hot components, together with the corresponding maximum
  uncertainties are listed in the rotation diagram panels.}
\label{fig:spectra}
\end{figure*}

\begin{figure*}[t]
\centering
\includegraphics[width=0.9\textwidth,angle=0]{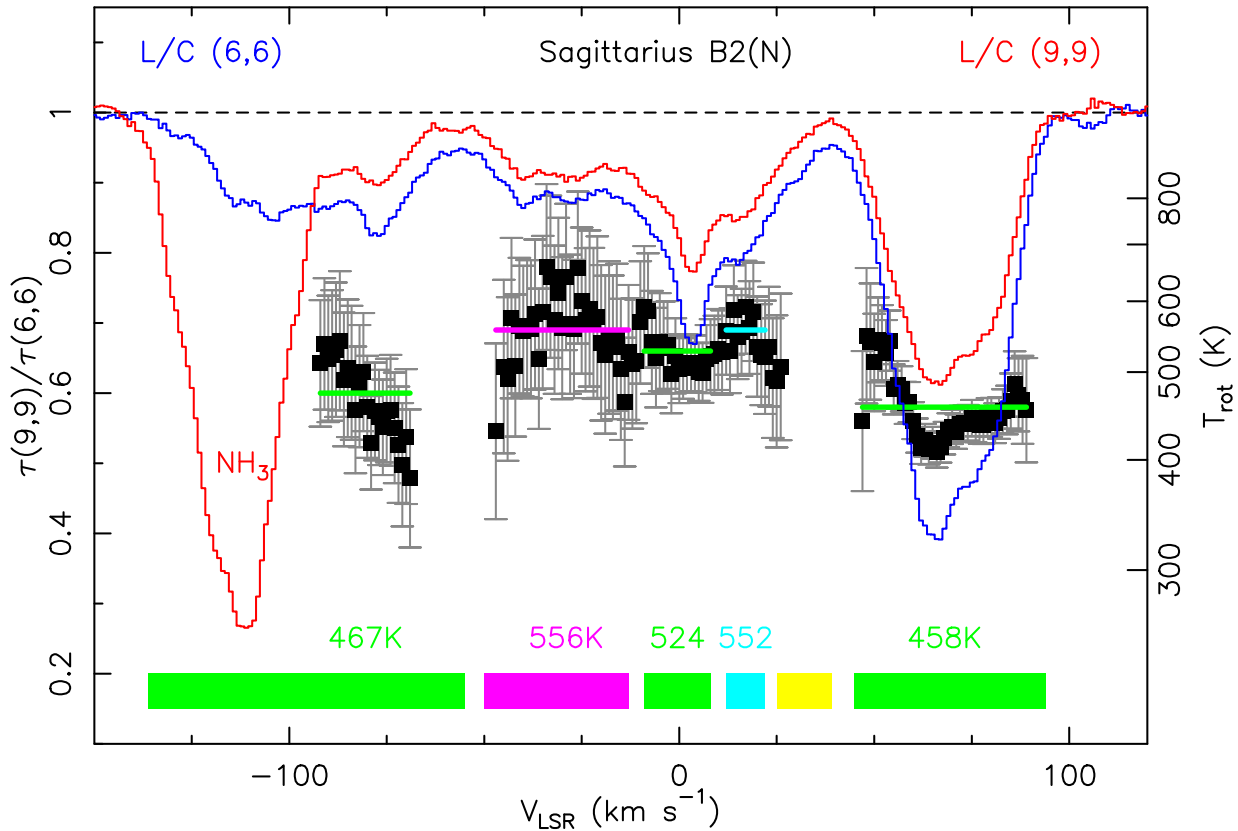}
\caption{Herschel/HIFI spectra of the (6,6) and (9,9) inversion
  transitions of the hydronium ion toward Sagittarius B2(N),
  normalized by the continuum, plotted as a function of the local
  standard of rest velocity (blue and red lines, respectively).
  Black points show the (9,9)/(6,6) optical depth ratio with the
  corresponding 1$\sigma$ uncertainty in a 1~\kms\ velocity bin (left
  vertical scale). The corresponding rotational temperature scale is
  shown on the right. Velocities corresponding to the Galactic center
  material, Norma, Scuttum, and Sagittarius arms are marked in green,
  magenta, cyan and yellow, respectively, above the horizontal axis
  (see Table~1). }
\label{fig:sgrb2}
\end{figure*}

\section{Sagittarius B2 Results} 

\begin{deluxetable*}{lcccccccc}
\tablecaption{\htop\ Spectroscopic Parameters and Absorption Line
      Parameters toward Sagittarius B2(M) \label{tab:fitm}}
\tablewidth{0pt}
\tablehead{
\colhead{Transition} & \colhead{$\nu_{ul}$} & \colhead{$A_{ul}$} &
\colhead{$g_u$} & \colhead{ $g_l$} & \colhead{$E_l$} & \colhead{$V$} &
\colhead{$\Delta V$} & \colhead{$\tau dv$}
}
\startdata
(1,1) & 1655.834 & $5.464\times10^{-2}$ & 6 & 6 & 0 & --- & --- &
---\\ 
$0_0^- - 1_0^+$ & 984.712 & $2.298\times10^{-2}$ & 4 & 12 & 7.3 &
$64.9\pm0.2 $& $13.4\pm0.5$ & $6.10\pm0.18$ \\
(2,2) & 1657.248 & $7.303\times10^{-2}$ & 10 & 10 & 42.7 &
$64.4\pm0.4$ & $12.9\pm0.9$ & $7.33\pm0.49$ \\
(3,3) & 1663.587 & $8.312\times 10^{-2}$& 28 & 28 & 103.1 &
$64.6\pm0.3$ & $14.7\pm0.7$ & $16.6\pm0.7$ \\
(4,4) & 1674.866 & $9.047\times10^{-2}$ & 18 & 18 & 181.2 &
$64.5\pm0.3$ & $15.6\pm0.8$ & $7.97\pm0.36$ \\
(5,5) & 1691.134 & $9.701\times10^{-2}$ & 22 & 22 & 276.9 &
$64.3\pm0.3$ & $15.6\pm0.7$ & $5.83\pm0.22$ \\
(6,6) & 1712.461 & $1.036\times10^{-1}$ & 52 & 52 & 390.2 &
$64.0\pm0.4$ & $15.7\pm0.7$ & $10.7\pm0.4$ \\
(7,7) & 1738.936 & $1.107\times10^{-1}$ & 30 & 30 & 521.1 &
$64.1\pm0.6$ & $15.1\pm1.2$ & $4.06\pm0.29$ \\
(8,8) & 1770.679 & $1.188\times10^{-1}$ & 34 & 34 & 669.5 &
$65.0\pm0.6$ & $17.4\pm1.5$ & $4.66\pm0.33$ \\
(9,9) & 1807.826 & $1.280\times10^{-1}$ & 76 & 76 & 835.3 &
$64.0\pm0.4$ & $16.2\pm1.0$ & $7.50\pm0.39$ \\
(10,10) & 1850.536 & $1.387\times 10^{-1}$ & 42 & 42 & 1019. &
$64.8\pm0.5$ & $16.8\pm1.2$ & $3.52\pm0.22$ \\
(11,11) & 1898.992 & $1.511\times10^{-1}$ & 46 & 46 & 1219. &
$62.9\pm1.3$ & $23.7\pm3.5$ & $1.94\pm0.23$ 
\enddata

\tablecomments{Entries in the table are: transition, frequency
  (GHz), spontaneous emission coefficient (s$^{-1}$), upper and lower
  level degeneracies, lower level energy (K), as well as the observed
  line center velocity, line width, and velocity integrated line
  opacity (\kms). The (1,1) line in the Galactic center sources is
  blended with the $2_{12}-1_{01}$ transition of H$_2^{18}$O and
  cannot be used in the analysis. }
\end{deluxetable*}

\begin{deluxetable*}{lccccccc}
  \tablecaption{\htop\ Absorption Line Parameters toward Sagittarius
      B2(N) \label{tab:fitn}}
\tablewidth{0pt}
\tablehead{
\colhead{Transition} &  \multicolumn{2}{c}{5~\kms} &
      \multicolumn{2}{c}{65~\kms} &
      \multicolumn{2}{c}{80~\kms} & \colhead{Total} \\
\colhead{} & \colhead{$V$} & \colhead{$\tau dv$} &
\colhead{$V$} & \colhead{$\tau dv$} & \colhead{$V$} &  \colhead{$\tau
  dv$} & \colhead{$\tau dv$}
}
\startdata
      $0_0^- - 1_0^+$ & $4.4\pm3.0$ & $1.59\pm0.30$ & $66.7\pm0.3$ & $9.7\pm0.3$
      & $81.9\pm0.7$ & $1.58\pm0.23$ & $12.9\pm0.5$ \\
      (2,2) & $4.1\pm1.9$ & $3.39\pm0.43$ & $66.0\pm 0.4$ &
      $13.0\pm0.4$ & $81.2\pm0.7$ &  $2.31\pm0.33$ & $18.7\pm0.7$ \\ 
      (3,3) & $6.3\pm1.3$ & $8.26\pm0.63$ & $66.0\pm0.3$ &
      $25.9\pm0.6$ & $80.6\pm0.5$ & $5.74\pm0.53$ & $40.\pm1.0$ \\
      (4,4) & $7.4\pm1.3$ & $3.10\pm0.25$ & $65.7\pm0.3$ &
      $11.3\pm0.3$ & $80.8\pm0.5$ & $2.87\pm0.21$ & $17.2\pm0.5$\\
      (5,5) & $7.6\pm2.0$ & $4.36\pm0.45$ & $64.3\pm0.6$ &
      $9.2\pm0.4$ & $80.2\pm0.7$ & $3.37\pm0.35$ & $17.0\pm0.7$\\
      (6,6) & $5.6\pm1.8$ & $8.85\pm0.90$ & $64.6\pm0.7$ &
      $18.8\pm0.9$ & $79.8\pm1.1$ & $4.45\pm0.76$ & $32.\pm1.5$\\
      (7,7) & $5.4\pm1.2$ & $3.39\pm0.27$ & $66.3\pm0.5$ &
      $7.4\pm0.3$ & $80.9\pm0.8$ & $1.85\pm0.23$ & $12.7\pm0.5$\\
      (8,8) & $3.2\pm2.2$ & $2.45\pm0.35$ & $65.2\pm0.8$ & $5.9\pm0.4$
      & $79.8\pm1.0$ & $1.83\pm0.31$ & $10.2\pm0.6$ \\
      (9,9) & $5.4\pm2.4$ & $4.94\pm0.75$ & $65.4\pm1.3$ & $9.4\pm0.8$
      & $79.1\pm1.4$ & $3.38\pm0.74$ & $17.8\pm1.3$\\
      (10,10) & $9.3\pm3.0$ & $1.62\pm0.35$ & $65.4\pm1.7$ &
      $3.35\pm0.39$ & $78.9\pm1.4$ & $1.51\pm0.36$ & $6.5\pm0.64$\\
      (11,11) & $3.7\pm3.2$ & $0.93\pm0.22$ & $64.1\pm1.4$ &
      $2.19\pm0.21$ & $79.8\pm1.4$ & $0.81\pm0.18$ & $3.9\pm0.35$
\enddata

\tablecomments{Entries in the table are: transition, line velocity
  and velocity integrated line opacity (\kms) for the three velocity
  components seen in the spectra, and total velocity integrated line
  opacity. Given the limited signal-to-noise
  ratio, the line widths of the three components
  have been fixed at 25.0, 18.8, and 10.6~\kms,
  respectively, as given by the fit to the averaged spectrum of all
  the transitions observed.}
\end{deluxetable*}

Continuum-divided spectra of the \htop\ inversion lines toward
Sagittarius B2(M) and (N) from the HEXOS spectral scans are shown in
Figure~\ref{fig:spectra}, upper and lower panels, respectively. The
lower panel is a revised version of Figure~3 of \cite{lis12},
including latest improvements in the data processing. The (1,1) line
is not included in the analysis, as it is blended with the strong
$2_{12}-1_{01}$ line of H$_2^{18}$O. Instead, the $J_K = 0_0^-
\leftarrow 1_0^+$ ground state ortho-\htop\ spectrum is included (see
energy level diagram in Figure~2 of \citealt{lis12}). Gaussian fit
results for the two sources are listed in Tables~\ref{tab:fitm} and
\ref{tab:fitn}. We use the latest line frequencies of \cite{yu14},
which in some cases differ by up to \about 25~MHz from those used in
\cite{lis12}. The \htop\ column densities in the metastable levels can
be derived directly from the velocity integrated optical depths of the
corresponding inversion transitions using the formula (see,
  e.g., eq. 3 of \citealt{neufeld10}):

$$ N_l({\rm \htop}) = { {8 \pi g_l} \over {A_{ul} g_u \lambda^3} } \int \tau dv~,$$

\noindent where $A_{ul}$ is the spontaneous emission coefficient,
$g_u$ and $g_l$ the upper and lower level degeneracies, and $\lambda$
the wavelength. Spectroscopic constants for the observed transitions
are included for completeness in Table~\ref{tab:fitm} and the
resulting rotation diagrams (plots of the logarithm of the integrated
line strength as a function of the lower level energy; e.g.,
\citealt{linke79}) for the two sources are also shown in
Figure~\ref{fig:spectra}. Toward Sagittarius~B2(M), the spectrum is
dominated by a single component centered around 64~\kms, with a line
width of 15~\kms, associated with the cloud envelope. Three velocity
components, centered around 5, 65, and 80~\kms, with line widths of
\about 25, 19, and 11~\kms\, respectively, can be identified in the
Sagittarius~B2(N) spectra. The strongest component at 65~\kms\ is
associated with the cloud envelope and contains \about 60\% of the
total \htop\ column density. The spectra are not sensitive enough to
produce independent rotational diagrams for the individual velocity
components. Therefore the three components have been added together to
produce a single rotational diagram for this source.\footnote{There is
  evidence that the 5~\kms\ component may have a somewhat higher
  rotational temperature than the other two, stronger components that
  are responsible for the bulk of the absorption (see
  Fig.~\ref{fig:sgrb2}).}

The rotational diagrams for the two sources naturally separate into
``warm'' and ``hot'' components (blue and red points in
Figure~\ref{fig:spectra}; lower level energies below and above 200~K,
respectively). Rotational temperatures, determined separately for the
warm and hot components from the least square fits to the data, are
shown in Figure~\ref{fig:spectra} with the corresponding
uncertainties, which should be viewed as \emph{maximum} uncertainties
due to systematics, such as baseline removal. We conservatively assume
a 25\% maximum uncertainty for the individual measurements.

The warm component toward the two sources is well described by a
rotational temperature of \about $175\pm50$~K, while the temperatures
of the hot component toward the two sources are consistent with
$485\pm65$~K (maximum uncertainties). A two-component
fit is only an approximation to the observed rotation diagram and a
continuous range of temperatures between these limiting values is
possible. However, one clear conclusion is that the population of the
highest and lowest-energy metastable levels is described by
\emph{different} rotational temperatures, \about 175 and 485~K,
respectively.

Total \htop\ column densities toward Sagittarius B2(N) and (M),
obtained by adding up the population in the observed rotational
levels, are $1.1 \times 10^{15}$ and $4.4 \times 10^{14}$~\pscm,
respectively, with \about 40\% of the total column density in the hot
component. The extended envelope of Sagittarius B2, with a velocity of
\about 65~\kms\ has a column density of $7 \times 10^{14}$~\pscm\ on
the line of sight toward Sagittarius B2(N). This is \about 50\% higher
than the corresponding value toward Sagittarius B2(M). This, together
with the fact that the 80~\kms\ component that is quite prominent in
the Sagittarius B2(N) spectra is not seen toward Sagittarius B2(M),
suggests that significant variations in the \htop\ column density are
present on linear scales of order 1.7~pc. If the absorption originated
in a foreground shocked screen between the Sun and Sagittarius~B2
(e.g., \citealt{ceccarelli02}), one would not expect such large column
density variations.

The new Sagittarius~B2(N) deep integrations are much more sensitive
than the HEXOS spectral scans. The continuum-divided spectra of the
(6,6) and (9,9) transitions are shown in Figure~\ref{fig:sgrb2} (blue
and red lines, respectively; average of the H and V polarizations,
equally weighted; left vertical scale). \htop\ absorption is clearly
detected over a wide range of LSR velocities between about --120 and
+90~\kms. The (9,9) absorption at the most negative velocities is
contaminated by the strong $J_K=3_1^- \leftarrow 2_1^+$ line of
ammonia in the same sideband, which may affect the derived line ratio
and the corresponding rotational temperature at these velocities. It
is not possible to remove the contamination, because the only other
para-\nhhh\ line with comparable line strength and lower level energy
is itself blended and the intrinsic line profile cannot be determined.

\begin{deluxetable}{lccc}
  \tablecaption{\htop\ (9,9)/(6,6) Line Ratios and Rotational
      Temperatures toward Sagittarius B2(N) \label{tab:trot}}
\tablewidth{0pt}
\tablehead{
\colhead{Component} & \colhead{$V$} & \colhead{$\tau(9,9)/$} &
\colhead{$T_{\rm rot}$} \\
\colhead{ }                   & (\kms) &  \colhead{$\tau(6,6)$}  & (K) 
}
\startdata
      Galactic center & --92 to --69 & $0.59\pm0.11$ & $467~(-80,+87)$\\
      Norma arm       & --47 to --13 & $0.69\pm0.11$ & $556~(-99,+124)$\\
      Galactic center & --9 to 8         & $0.66\pm0.08$ & $524~(-71,+85)$\\
      Scutum arm     & 12 to 22          & $0.69\pm0.12$ & $552~(-107,+141)$\\
      Sagittarius B2   & 47 to 89         & $0.58\pm0.04$ & $458\pm31$
\enddata

\tablecomments{Uncertainties are conservative maximum estimates, allowing a
factor of 2 correction due to correlated noise.}
\end{deluxetable}

Black points in Figure~\ref{fig:sgrb2} show the (9,9)/(6,6) optical
depth ratio in 1~\kms\ channels (left vertical scale). Only points
corresponding to velocities at which absorption in both transitions
is detected above $4\sigma$ level in individual channels are plotted. The
corresponding $1\sigma$ uncertainties are computed from the difference
between the H and V polarization spectra, which are assumed to be
independent measurements. The rotational temperature describing the
population of these high-energy ($>$400~K above the ground rotational
state) metastable levels is uniformly high, in the range 400--700 K
(Figure \ref{fig:sgrb2}; right vertical scale). 

The differential rotation of the Milky Way allows, in principle,
separating in velocity space spectral features from gas clouds at
different galactocentric radii (see, e.g., \citealt{vallee08},
however, the distance determination and hence the assignment to
specific spiral arms is complicated due to the streaming motions in
the arms, \citealt{reid09}). All velocity components that can be
clearly identified in the Sagittarius B2(N) and (M) spectral scan data
correspond to the Galactic center gas. There is evidence for broad
absorption in the (3,3) and (6,6) lines toward Sagittarius~B2(N), but
the signal-to-noise ratio is limited. However, \htop\ absorption is
clearly detected over a broad range of velocities in the deep
integrations toward this source (Figure~\ref{fig:sgrb2}). Velocities
of the various foreground components on this line of sight are marked
with horizontal color bars above the abscissa: green--Galactic center
gas, magenta--Norma arm, cyan--Scutum arm, and yellow--Sagittarius
arm. Absorption at velocities of the Norma and Scutum arms (magenta
and cyan, respectively) is clearly detected, in addition to the
Galactic center gas. Averaged optical depth ratios and rotational
temperatures for five velocity ranges corresponding to different
kinematic components are listed in Table~\ref{tab:trot}.

Determining the uncertainty of the velocity-averaged rotational
temperature values is not trivial, as the channels are not independent
due to the presence of baseline ripples in the spectra. This aspect
has been discussed in detail in the context of the measurements of the
water ortho/para ratio by \cite{lis13}. Based on our previous
experience with HIFI HEB data, we allow a factor of 2 correction for
the uncertainty of the average line ratio to account for the
correlation between the spectral channels. This leads to conservative
maximum uncertainties for the line ratio and the rotational
temperature listed in Table~\ref{tab:trot}. We conclude that, within
the observational uncertainties, the rotational temperature of the hot
component is the same, \about 500~K, in regions at different
galactocentric distances. As a comparison, the kinetic temperature in
the foreground clouds, determined from the \nhhh\ (1,1) and (2,2)
observations, is in the range 10--60~K \citep{tieftrunk94}. These
low-evergy transitions, however, are not sensitive to the presence of
high-temperature gas.

\section{W31C Results}  

\begin{deluxetable}{lc}
\tabletypesize{\scriptsize}
\tablecaption{\htop\ Absorption Line Strength toward W31C \label{tab:fitw}}
\tablewidth{0pt}
\tablehead{
\colhead{Transition}  & \colhead{$\tau dv$}
}
(1,1) & $0.30\pm0.04$ \\
$0_0^- - 1_0^+$ & $0.30\pm0.04$ \\
(2,2) & $0.30\pm0.05$ \\
(3,3) & $0.40\pm0.04$ \\
(6,6) & $0.17\pm0.02$ \\
(9,9) & $0.08\pm0.02$ \\
\tablecomments{Entries in the table are: transition and velocity
  integrated line opacity (\kms).}
\end{deluxetable}

\begin{figure}[t]
\centering
\includegraphics[width=0.95\columnwidth,angle=0]{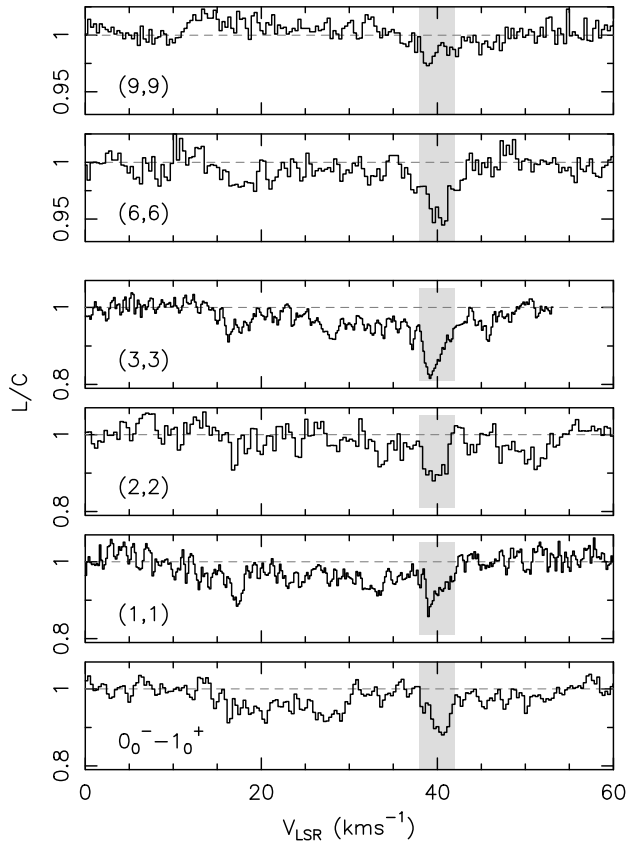}
\caption{Spectra of the (9,9) and (6,6) inversions transitions of the
  hydronium ion absorption toward W31C, combined with earlier
  observations of lower-energy transitions obtained within the PRISMAS
  GT KP. Weak \htop\ absorption is detected at \about 40~\kms\
  (highlighted in light gray). Due to the narrow line width, the (1,1)
  absorption in this source is not blended with H$_2^{18}$O and this
  transition can be used in the analysis. Note that the vertical scale
  is different for the (6,6) and (9,9) lines. }
\label{fig:w31c}
\end{figure}

\begin{figure}[t]
\centering
\includegraphics[width=0.95\columnwidth,angle=0]{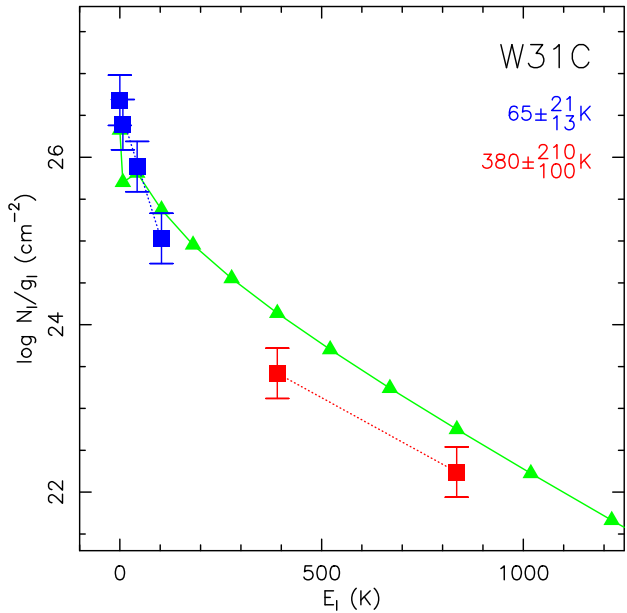}
\caption{Observed population diagram for \htop\ toward W31C (blue and
  red squares). Derived rotational temperatures of the warm and hot
  component are given together with the corresponding maximum
  uncertainties, computed assuming 35\% maximum uncertainties of the
  individual measurement. Green triangles show the \htop\ population
  of the metastable levels computed using the simple formation model
  described in the text. The model is included for illustrative
  purposes only and is not meant to be a fit to the data.}
\label{fig:pop}
\end{figure}

Gas motions in the Galactic center region are quite complex, since the
velocity field is dominated by the non-axisymmetric bar potential. In
particular the gas in the so-called ``x2'' non-circular orbits may
overlap in velocity with the foreground disk clouds. One thus cannot
exclude that the \htop\ absorption at \emph{all} velocities toward
Sagittarius B2 is predominantly associated with the gas in the Central
Molecular Zone. The observations of W31C, a line of sight that does
not intersect the Galactic center region, provides independent
constraints. The (6,6) and (9,9) excited inversion lines of the
hydronium ion are also detected in the deep HIFI observations on this
line of sight (Figure~\ref{fig:w31c}), which samples the diffuse
interstellar medium along a 4~kpc line of sight through the Galactic
disk. The foreground absorption covers velocities between 10 and
50~\kms\ \citep{gerin10, neufeld10}, with a strong feature at \about
40~\kms\ clearly seen in the \htop\ spectra. The observed absorption
line intensities are listed in Table~\ref{tab:fitw}. The resulting
rotational diagram (Figure~\ref{fig:pop}) can again be described by
two components with temperatures of 65~K (+21,--13)~K and 380~K
(+210,--100)~K. The uncertainties are maximum uncertainties, assuming
a 35\% maximum uncertainty for the individual measurements (a
conservative estimate of the combined statistical and systematic
uncertainties). Although the uncertainty in the rotational temperature
of the hot component is large, the lower bound is well constrained,
demonstrating that the population of the excited metastable levels is
described by a temperature that is significantly higher than that
describing the low-energy metastable rotational levels, in agreement
with the Sagittarius B2 results. The 65~K temperature describing the
population of the low-energy metastable levels is consistent with the
kinetic temperature of the gas in the diffuse interstellar medium in
the Galactic disk \citep{snow06}.

From the two-component fit to the rotation diagram, we estimate a
total \htop\ column density of $1.2 \times 10^{13}$~\pscm\ toward
W31C, with \about 20\% in the hot component, as compared to \about
40\% toward Sagittarius B2(N) and (M).

\section{Discussion}

Earlier ground-based \citep{hutte95} and space-borne
\citep{ceccarelli02} observations of ammonia absorption toward
Sagittarius B2 have been interpreted in terms of a physically hot
component of molecular gas in the central region of the Galaxy, in
which the observed high rotational excitation temperature traces a
correspondingly high kinetic temperature. At first glance this seems
plausible, because shock waves in this active high-mass star-forming
region should be able to maintain some molecular gas at temperatures
of 500 to 700~K, or because the region contains luminous heating
sources of X-rays and cosmic rays. Moreover, strong X-ray emission in
iron fluorescence at 6.4~keV has been taken as evidence that the
envelope of Sagittarius B2 was recently illuminated by an X-ray flash,
attributable to a flare from the region around the central black hole
that is associated with the Sagittarius A$^*$ radio continuum source
\citep{sunyaev93, koyama96}. The Sagittarius B2 X-ray flux is now
fading \citep{terrier10}, with a characteristic decay time of 8
years, comparable to the light-travel time across the central part of
the molecular cloud.

While the competing mechanisms put forward to explain the hot gas
component on the line of sight toward Sagittarius B2 are all
plausible, a generally accepted explanation has yet to be offered.
Interestingly, recent observations of the (8,8) to (15,15) ammonia
inversion lines \citep{mills13} indicate rotational temperatures of
350--450 K in several largely \emph{quiescent} Galactic center giant
molecular clouds. This indicates that the effect is not limited to
Sagittarius B2, but widespread in the Galactic center environment. The
hot gas component traced by the high-energy ammonia inversion lines
would thus have to fill a few hundred parsec size region. The energy
input required to heat such a vast volume of gas would then have to be
carefully evaluated. Moreover, the diffuse, widespread, low-density
gas component in the Central Molecular Zone can be independently
probed through the infrared absorption spectroscopy of the H$_3^+$
rovibrational transition \citep{oka05}, which indicate temperatures of
``only'' \about 250~K. Inversion lines of ammonia, up to (10,10), have
also been detected in absorption toward PKS~1830-211, indicating
rotational temperatures \ga 600~K for the highest-energy lines
\citep{henkel08}. The absorption is attributed to molecular gas in
spiral arms of an ordinary galaxy at a redshift of 0.89.

Explanations similar to those previously invoked to explain the
high-rotational temperatures of ammonia cannot easily be applied to
H$_3$O$^+$: if the density were high enough to thermalize the
populations of highly excited rotational states at a kinetic
temperature of 400 to 600~K, then molecular collisions would also
raise the excitation temperatures of the high-$J$ inversion doublets
to values higher than the brightness temperature of the continuum
radiation at $\lambda 150 - 200$~\mic. In that case, the inversion
lines could not appear in absorption as is observed. A fundamentally
different interpretation is required to explain simultaneously the
high rotational excitation of H$_3$O$^+$, its short chemical lifetime,
and its relatively large total column density, $N({\rm H}_3{\rm O}^+)
= 4-7\times 10^{14}$ cm$^{-2}$, close to an intense far-infrared
continuum source. 

The characteristic two-component rotational diagram for metastable
states of H$_3$O$^+$ naturally results from its formation by exoergic
ion-neutral reactions. The excess enthalpy (heat of formation) of the
main source reaction (e.g., \citealt{herbst73})
$$ {\rm H}_2{\rm O}^+ + {\rm H}_2 \to {\rm H}_3{\rm O}^+ + {\rm H} $$
goes partly into internal rotational excitation of the product ion,
the highly excited non-metastable states relax rapidly by spontaneous
radiative transitions to the metastable states (i.e. the lower $J=K$
inversion levels), whose populations are limited by the rates of
excited-molecule formation and chemical destruction. Populations are
partly re-distributed by inelastic collisions and by absorption of the
background far-infrared continuum radiation. The populations of the
lowest metastable states may approach a collision-dominated thermal
distribution. The turnover to a ``hot'' formation-dominated
distribution is detectable when the formation rate is a
non-negligible fraction of the inelastic collision rates.

The relation between molecular excitation and formation can be
outlined quantitatively as follows. The destruction rate per
H$_3$O$^+$ molecule by dissociative recombination with electrons can
be written ${\cal D} = n(e) k_{\rm dr}$ s$^{-1}$, where the rate
coefficient for the process $k_{\rm dr} = 4.3\times 10^{-7}
(T/300)^{-1/2}$ cm$^3$ s$^{-1}$ at kinetic temperature $T$
\citep{jensen00} and $n(e)$ is the number density of free electrons.
At $n(e)\sim 0.01$ to $0.1$ cm$^{-3}$, and $T\sim 50$ to $200$ K,
${\cal D} \approx 10^{-8}$ to $10^{-7}$ s$^{-1}$. This is fast enough
to be the dominant destruction mechanism. If we furter assume steady
state between destruction and formation, then the formation rate per
unit voulme can be written as ${\cal F} = n({\rm H}_3{\rm O}^+) {\cal
  D}$~cm$^{-3}$s$^{-1}$. With no further knowledge of the chemistry,
we can assert that the source of H$_3$O$^+$ in the envelope of
Sagittarius B2 must be of the order of
$$ {\cal F} = N({\rm H}_3{\rm O}^+) {\cal D} / L \approx 10^{-11}
\Bigl[{{1 \;{\rm pc}}\over{L}}\Bigr] \Bigl[{{n(e)}\over{0.1\;{\rm
      cm}^{-3}}}\Bigr] \;{\rm cm}^{-3}{\rm s}^{-1}, $$
where $L$ is the characteristic length scale of the absorbing region.
The interstellar chemistry of oxygen-containing ions is thought to be
straightforward \citep{herbst73}, with cosmic-ray or X-ray ionizations
of hydrogen leading to H$^+$ and H$_3^+$, which transfer charge to
O$^+$ or OH$^+$. The oxygen ions form H$_2$O$^+$ and finally
H$_3$O$^+$ via reactions with H$_2$. In regions where the atomic
fraction is small, H/H$_2 \ll 1$, the chemical source rate of the
terminal ion H$_3$O$^+$ can be a significant fraction of the hydrogen
ionization rate. Thus we can express ${\cal F} \sim \zeta n_{\rm H}
\epsilon$, where $\zeta$ is the ionization rate, $n_{\rm H}$ is the
density of hydrogen, and $\epsilon$ is a chemical loss factor that
accounts for minor channels in the ion chemistry that prevent the
hydrogen ions from processing solely the oxygen. Such an analysis has
been applied to Herschel/HIFI observations of OH$^+$ and H$_2$O$^+$ in
diffuse molecular gas by \cite{neufeld10}, who concluded that those
highly reactive ions arise mostly in components of gas with a high
atomic content, unlike \htop, which traces predominantly molecular
gas. Following these arguments, we see that the observed column
density and estimated destruction rate of H$_3$O$^+$ toward
Sagittarius B2 require
$$ \zeta n_{\rm H} \leq 10^{-11}/\epsilon \;\;\;{\rm cm}^{-3}\;{\rm s}^{-1} \;,$$
for $L\approx 1$ pc and $n(e)\leq 0.1$ cm$^{-1}$ in the envelope of
Sagittarius B2. This could be achieved with a cosmic-ray ionization rate
$\zeta \sim 10^{-15}$ s$^{-1}$ at $n_{\rm H}\epsilon \sim 10^4$
cm$^{-3}$, or with less extreme values if the electron density is
lower. Cosmic ray ionization rates in excess of $10^{-15}$ s$^{-1}$
have been deduced from infrared absorption observations of H$_3^+$ on
several lines of sight through the Central Molecular Zone of the
Galaxy (\citealt{gotto13}, and references therein).

The line of sight to W31C presents a simpler test case because it
lies entirely outside the Central Molecular Zone, where conditions are
extreme. For illustration, consider the low-density limit, in which
collisional excitation is neglected and the rotational populations are
governed entirely by the formation process and radiative transitions.
Let the state-specific formation rate of H$_3$O$^+$ be described by
$$  F(J,K,p) = {\cal F} g(J,K,p) \exp\bigl(-E(J,K,p)/kT_f \bigr) / Q(T_f)  $$
for each state of rotational quantum numbers $J$, $K$, and parity $p$,
where $g$ is the statistical weight, $E(J,K,p)$ is the energy, $Q$ is
the partition function, and $T_f$ is a parameter---the formation
temperature. The populations of the vibration-rotation levels are then
computed by solving the rate equations with formation, destruction,
and all radiative transitions included (equations 12--15 of
\citealt{vandertak07}), but with all the collisional terms set to
zero. The results of one example are compared with the observed
population diagram in Figure~4. In this case, ${\cal D}=10^{-7}$ s$^{-1}$
(corresponding to $n(e)\approx 0.1$ cm$^{-3}$ at $T<100$ K) and the
total column density is $N({\rm H}_3{\rm O}^+) = 1.4\times 10^{13}$
cm$^{-2}$. The formation temperature is $T_f = 400$ K and the
molecules are exposed to an average Galactic background continuum
similar to that described by \cite{black94}. At the adopted destruction
rate, ${\cal F} = 4.5\times 10^{-14}$ cm$^{-3}$ s$^{-1}$ is required
to attain a good match in column density over a total path length
$L\approx 10$ pc. This rate, in turn, would imply a rate of cosmic-ray
ionization of H$_2$, $\zeta_2 = {\cal F}/n({\rm H}_2)\epsilon \approx
5\times 10^{-16}$ s$^{-1}$ for a density $n({\rm H}_2) = 300$
cm$^{-2}$ and $\epsilon\approx 0.3$. This ionization rate would be
consistent with that derived by \cite{indriolo12} toward W51
IRS2 if the H$_3$O$^+$ is found preferentially in regions of higher
molecular fraction and higher $\epsilon$ than the observed OH$^+$.
These conditions are also in harmony with photochemical models of
diffuse molecular gas as discussed by \cite{hollenbach12}.

A combined excitation-chemistry model, which includes explicitly the
exothermicity of the chemical reactions leading to the formation of
the hydronium ion and ammonia is beyond the scope of the present paper
and will be presented separately (J. Black, in prep.) Such
calculations are clearly needed to demonstrate that chemistry and
excitation are closely coupled in some interstellar environments and
should not be treated separately and to use \htop\ as a quantitative
tracer of the cosmic ray ionization rate\footnote{One serious
  impediment is that the collisional cross-sections for \htop\ are
  only available for rotational levels with energies below \about
  350~K, scaled from ammonia \citep{offer92}.}. This aspect is of
particular interest for understanding the energetics of the nuclear
regions of ultraluminous, star-forming galaxies. Highly-excited
inversion lines of the hydronium ion have now been detected in the
ultraluminous infrared galaxy Arp~220, as well as in NGC~4418, which
harbors a deeply buried, compact AGN \citep{gonzalez13}. In these
environments, the hydronium ion absorption traces a relatively
low-density molecular gas with a large filling factor, which is
directly influenced by the feedback process of the central engine.

The traditional explanation for these observations invokes a new phase
of the interstellar medium, which would have to be heated by some
combinations of X-rays, cosmic rays, shocks, or turbulence to \about
500 K. If the chemical formation pumping explanation is correct, this
additional energy input is not required---the observations can simply
be explained by chemical formation pumping in a much lower temperature
medium. This relaxes the heating requirements and simplifies the
physical picture of the interstellar medium in these active
environments. With the improved understanding of the excitation, the
far-infrared inversion lines of the hydronium ion also provide an
accurate measurement of the ionization rate in the mostly molecular
gas component of high-redshift galaxies, using ALMA, highly
complementary to the OH$^+$ and H$_2$O$^+$ observations that primarily
trace atomic gas, with low molecular fraction \citep{gerin10,
  neufeld10}.

Symmetric top molecules have been proven to be some of the best
tracers of the gas kinetic temperature in the interstellar medium
\citep{ho83, mangum13}. However, the results presented here suggest
that they may sometimes fail as an interstellar thermometer, under
specific conditions. This typically involves low-density, diffuse gas,
as collisions and strong radiation field tend to drive the rotational
level population to equilibrium conditions. Consequently, care has to
be taken when interpreting the derived rotational temperatures as
physical temperatures of the medium.

%

\acknowledgments HIFI has been designed and built by a consortium of
institutes and university departments from across Europe, Canada and
the United States (NASA) under the leadership of SRON, Netherlands
Institute for Space Research, Groningen, The Netherlands, and with
major contributions from Germany, France and the US. Support for this
work was provided by NASA (\emph{Herschel} OT funding) through an
award issued by JPL/Caltech. The research of PS, CC, and RH is
supported by the Collaborative Research Center 956 funded by the
Deutsche Forschungsgemeinschaft (DFG).

\end{document}